\documentclass[aps,prl,reprint,showpacs,superscriptaddress]{revtex4-1}
\usepackage{epsfig}
\usepackage[]{natbib}
\usepackage{xspace}
\usepackage{color}
\usepackage{bbold}
\usepackage{tabu}
\usepackage{mathtools}

\begin{document}

\title{
Quantum Renewal Equation for the first detection time of a quantum walk 
 }

\author{H. Friedman}
\author{D. A. Kessler}
\author{E. Barkai}
\affiliation{Department of Physics, Institute of Nanotechnology and Advanced Materials, Bar Ilan University, Ramat-Gan
52900, Israel}

\begin{abstract}
We investigate the statistics of the first detected passage time
of a quantum walk. 
The postulates of quantum theory, in particular the collapse of
the wave function upon measurement,
reveal an intimate connection between the wave function of 
a process free of measurements, i.e. the solution of the Schr\"odinger
equation, and the statistics of first detection events on a site.
For stroboscopic measurements a quantum renewal equation yields
basic properties of quantum walks. For example, for a tight binding
model on a ring we discover critical sampling times, diverging
quantities such as the mean time for first detection,
and an optimal detection rate. For a quantum
walk on an infinite line
the probability of first detection decays like $(\mbox{time})^{-3}$ with 
 a superimposed
oscillation, critical behavior for a specific choice of sampling time,
 and vanishing amplitude when the sampling time
approaches zero  due to the quantum Zeno effect. 
\end{abstract}
\maketitle

A century ago Schr\"odinger \cite{Schro} 
discovered a fundamental  connection between occupation
probabilities and first passage time distributions \cite{MontrollW,Redner}, 
a relationship used to derive some of  the most basic properties
of classical random walks.
Ever since, the problem of first passage time statistics has attracted
tremendous interest, as it is applicable to many fields of science
\cite{MOR,Benichou}. 
In a nutshell the main idea  \cite{Schro} is simple: 
the path of a Markovian random walk on a graph, starting
at $x$ and reaching $0$ at time $t$, can be decomposed into a path that 
arrived at $0$ for the {\em first} time at time $t'$, 
and then an {\em independent} segment
which starts at $0$ and  returns
back in the time  interval $t-t'$.   
The consequence is a well known renewal formula, the first
equation in Redner's monograph \cite{Redner}, 
relating  occupation probabilities
and first passage time statistics.  
More recently, quantum walks have attracted much interest both theoretically
\cite{Aharonov1,Ambainis,Blumen}
and experimentally 
\cite{Yaron,QWe2,Xue}.
These exhibit interference patterns and ballistic scaling and in that
sense exhibit behaviors drastically different from the classical random
walk. 
Particularly controversial has been the question of  the
first passage time  for quantum dynamics.
The latter is ill defined,
so  we consider 
 the first {\em  detected} passage time to  a site
(see details below)  
\cite{Lumpkin,Bach,Stefanak,Das,Miquel,Krap,Dhar,Dhar1}.
Our main result is the quantum analogue of 
Schr\"odinger's classical renewal equation. With this equation,
which deals with amplitudes,  rather than probabilities, we calculate
 some
basic properties of quantum walks, for example the first {\em detected}
passage time
statistics on a line. Our approach is based on a projective
measurement scheme recently introduced by Dhar et al. \cite{Dhar,Dhar1}. 
We note that the quest for the quantum renewal equation, the question
of first passage time in quantum mechanics, or more generally 
the fluctuations of time observables, has a long hotly debated
history.
Here we use the textbook postulates
of quantum measurements, in particular the projective postulate 
\cite{CT}, to address the first detection problem of 
quantum dynamics, the main restriction being that
 the Hamiltonian is time independent,
the latter  corresponding to
the  Markov assumption
used by Schr\"odinger in the classical domain. 

{\em Model and the measurement process.} 
We consider a single quantum particle
 on a  graph, for example a 
lattice or a discretized ring,
 whose state
wave function is $|\psi\rangle$.
Under stroboscopic observations at times
$\tau,2 \tau, \cdots$. 
an observer performs measurements at a spatial position
which we may call $x=0$  
which is represented as the vector $|0\rangle$ (see schematic diagram in Fig.
\ref{figBenzene}).
Such stroboscopic measurements are  useful as they capture
quantum periodicities as shown below. 
A measurement provides two possible outcomes:
either the particle is detected at $x=0$  or it is not. 
The experiment provides the string: no, no, no, $\cdots$
 and finally at the
$n$-th attempt a yes so that 
$n \tau$ is the first detected passage time, whose statistics are investigated.
For that we must define the measurement process precisely \cite{Dhar,Dhar1}.

Just prior to the first measurement the wave function is 
$|\psi(\tau^{-}) \rangle = U(\tau) |\psi(0)\rangle$
and $U(\tau)= \exp( - i H \tau )$ is the unitary  evolution
operator,  $H$ the time independent  Hamiltonian,
$|\psi(0)\rangle$ is the wave function at the initial time $t=0$ and $\hbar=1$. 
For example, we will later investigate
the tight-binding model
\begin{equation}
H = - \gamma \sum_{x=-\infty}^\infty \left(|x\rangle \langle x+1 | + 
|x+1  \rangle \langle x| \right).
\label{eq01}
\end{equation}
This  describes a quantum particle jumping between nearest neighbours
on an infinite  one dimensional lattice so $x$ is an integer and
in that sense the model describes a quantum random walk \cite{Blumen}. We stress that
our main results are not limited to a specific Hamiltonian. 

 The probability of finding the
particle in state  $|0\rangle$, at the first measurement, is
$P_1 = |\langle 0 | \psi(\tau^{-})\rangle|^2$. 
If the outcome of the first  measurement is positive we get $n=1$.
 On the other hand, if the particle
is not detected, with probability $1-P_1$,
Von Neumann's postulate of collapse
 states that the  null
measurement alters the wave function in such a way
that the probability of detecting the
particle at the detection point $x=0$ at time $\tau^{+}\equiv \tau + 0^{+}$
 is zero.
 Afterwards, the evolution of the quantum state will
resume until the next measurement time $2 \tau$ 
via the transformation $U(\tau)$. 
In this sense we are considering projective measurements whose  duration
is very short, while between the measurements 
the evolution is according to the Schr\"odinger equation. 

\begin{figure}
\centering
\includegraphics[width=0.32\textwidth]{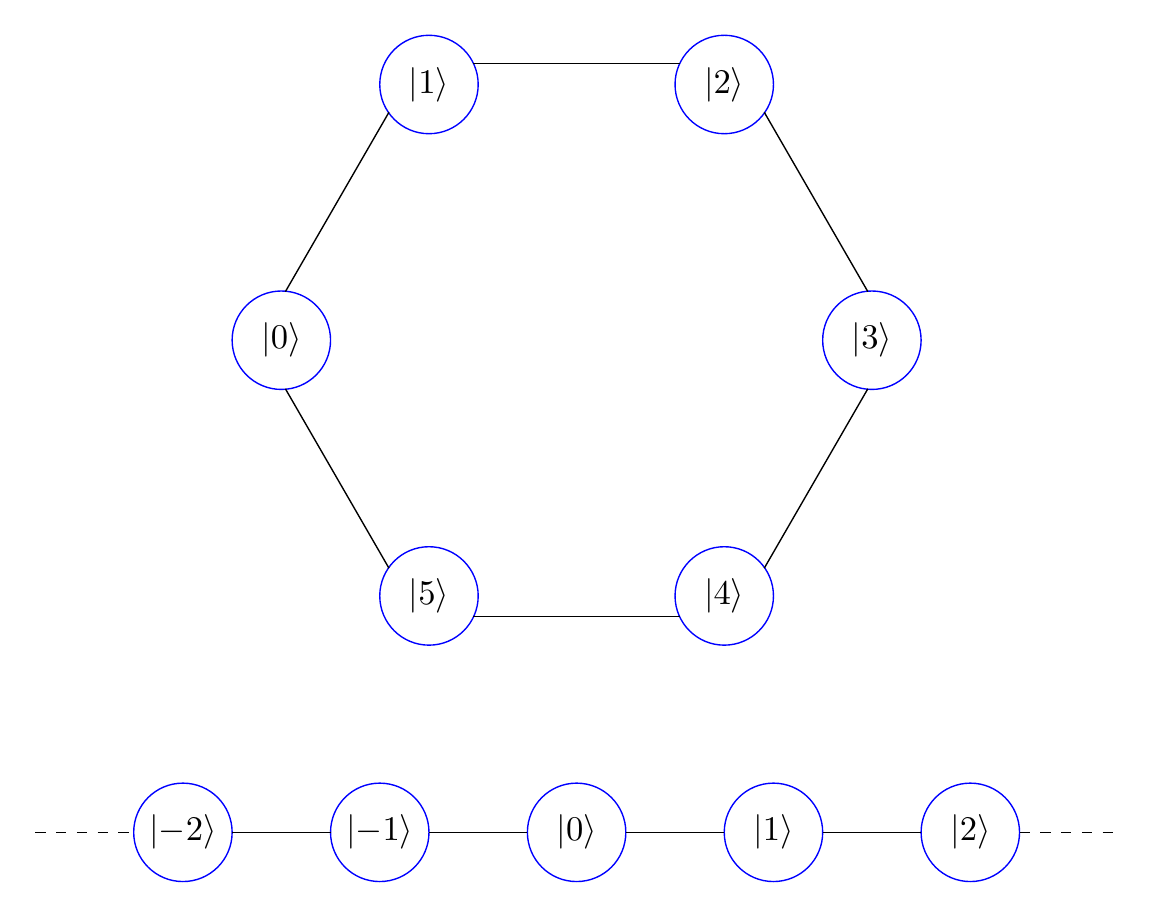}
\caption{
Schematic models of a benzene-like ring, representing a closed
geometry,  and an infinite one dimensional
lattice. 
Measurements on site $x=0$ are made
stroboscopically at  time $\tau,2\tau, ...$ to capture periodicities
in the underlying dynamics.
}
\label{figBenzene}
\end{figure}

Since the outcome of a null measurement is zero amplitude for 
finding the particle at $x=0$ 
at time $\tau^{+}$
 we have
\begin{equation}
| \psi(\tau^{+}) \rangle = N \left( \mathbb{1} - | 0 \rangle \langle 0 |\right)| \psi(\tau^{-}) \rangle,
\label{eq03}
\end{equation}
where $\mathbb{1}$ is the identity operator, 
and $N$ is determined from the normalization condition.
Since just prior to the
measurement the probability 
of finding the particle at $x\neq 0$ is $1-P_1$,
we get \cite{CT}
\begin{equation}
| \psi(\tau^{+}) \rangle = 
{ \mathbb{ 1} -  \hat{D}  \over \sqrt{1 - P_1}} U(\tau) |\psi(0)\rangle.
\label{eq04}
\end{equation}
where $\hat{D}=|0\rangle\langle 0|$ is the measurement's projection operator.
The probability
of detecting the particle at the second measurement,
conditioned that the quantum walker was not found in the first attempt, is
$P_2= | \langle 0 | U(\tau)| \psi(\tau^{+})\rangle|^2$.
This procedure  is continued to find the
probability of first detection in the $n$-th measurement, conditioned that
prior measurements did not detect the particle
 \cite{Dhar,Dhar1}, 
\begin{equation}
P_n = {\bigg| \langle 0 | \left[U(\tau) (1 - \hat{D} )\right]^{n-1} U(\tau) | \psi(0)\rangle\bigg|^2 \over (1 - P_1 ) .... (1- P_{n-1}) }. 
\label{eq08}
\end{equation}
In the numerator (respectively, the denominator) the operator 
 $1 - \hat{D}$ (the probabilities of non-detection $1-P_j$)  is found $n-1$ times corresponding
 to the $n-1$ prior measurements. 

%
%

{\em First detection probability $F_n$.}
The main focus of this work is on the probability of first detection
in the  $n$-th attempt, denoted $F_n$. 
This detection
consists of a set of $n-1$ null measurements, each 
weighed  by the conditional probability $1-P_j$, followed by
a positive measurement at attempt $n$, giving
\begin{equation}
F_n = (1-P_1)(1-P_2)... (1 - P_{n-1}) P_n.
\label{eqMult}
\end{equation}
Using Eq. (\ref{eq08}),
%
$F_n = |\phi_n|^2$
%
where 
\begin{equation}
\phi_n = \langle 0| U(\tau) \left[ \left( 1 - \hat{D} \right) U\left( \tau \right) \right]^{n-1} |\psi(0)\rangle
\label{eq09}
\end{equation}
is the first detection amplitude. 

{\em Solution using generating functions.}
 Eq. (\ref{eq09}) gives  $\phi_1= \langle 0| U(\tau) | \psi(0)\rangle$, 
$\phi_2= \langle 0 | U( 2 \tau)|\psi(0)\rangle - \phi_1 \langle 0 | U(\tau) | 0\rangle$ 
and by induction we find 
\begin{equation}
\phi_n = \langle 0 | U(n \tau) | \psi(0)\rangle -
\sum_{j=1} ^{n-1} \phi_j \langle 0 | U\left[\left(n-j\right) \tau\right] | 0 \rangle.
\label{eq15}
\end{equation}
This iteration rule 
yields the amplitude $\phi_n$ in terms 
of a propagation free of measurement, i.e. 
$ \langle 0 | U(n \tau) | \psi(0)\rangle$
 is the amplitude for being at the origin at time $n \tau$ in the absence of measurements, from which we
subtract $n-1$ terms related to the previous null measurements of the particle.
In practice, it is more convenient to work in terms of the generating
function \cite{Brown} 
also called the $Z$ transform
$\hat{\phi}(z) \equiv \sum_{n=1}^\infty z^n  \phi_n$. 
Multiplying  Eq. (\ref{eq15}) by $z^n$ and summing over $n$
using the convolution theorem we get
\begin{equation}
\hat{\phi}(z) = { \langle 0 | \hat{U}(z) | \psi(0)\rangle \over 1 + \langle 0 | \hat{U}(z) |0 \rangle},
\label{eq21}
\end{equation}
where $\hat{U}(z) \equiv \sum_{n=1} ^\infty z^n U(n \tau)$. 
From the generating function we obtain useful information on the
process using inversion formulae provided in the supplementary material (SM). 

{\em Quantum renewal equation.} 
For a particle free of any measurement, 
the amplitude of being at the
origin at time $t$  is
 $\langle 0|\psi_f(t)\rangle$ and 
$|\psi_f(t)\rangle = \exp( - i H t) |\psi_f(0)\rangle$.
Here  $|\psi_f(t)\rangle$ is 
the solution of
the Schr\"odinger equation  $i \partial_t |\psi_f\rangle = H |\psi_f\rangle$,
with the same initial conditions as for the first detection problem
under investigation $|\psi_f(0)\rangle=|\psi(0)\rangle$. 
Let us consider the initial condition where the particle is initially
localized at $x=0$ and  so $|\psi(0)\rangle = |0\rangle$.
 Using $\langle 0|0 \rangle = 1$,  
Eq. (\ref{eq21}) is rewritten
\begin{equation}
\hat{\phi}(z) = 1 - {1 \over \langle 0  | {1 \over 1 - z e^{-i H \tau}} | 0 \rangle}.
\label{eqRed03}
\end{equation}
We define the measurement-free generating function 
\begin{equation}
\langle 0 | \psi_f(z)\rangle_0  \equiv  \sum_{n=0} ^\infty z^n \langle 0 | \psi_f( n \tau)\rangle,
\label{eqRed05}
\end{equation}
and clearly $\langle 0 | \psi_f(z)\rangle_0 =\sum_{n=0} ^\infty \langle 0 | z^n \exp( - i H \tau n) | 0 \rangle$, the subscript zero 
denoting  the initial condition. 
Summing  the geometric series,  we get the appealing result
\begin{equation}
\hat{\phi}(z) = 1 - {1 \over \langle 0 | \psi_f(z)\rangle_0} .
\label{eqRed06}
\end{equation}
Thus, the generating function of the
 first detection time amplitude $\hat{\phi}(z)$
is determined from the $Z$ transform
of the wave function at the point of detection $x=0$.

 Similarly, for an initial condition initially localized at a site $x\neq 0$,
so that $|\psi(0)\rangle=|x\rangle$, for detection at site $0$ we find
that
\begin{equation}
\hat{\phi}(z) = {\langle 0 |\psi_f(z)\rangle_x \over \langle 0 | \psi_f(z)\rangle_0},
\label{eqRed06aa}
\end{equation}
where $|\psi_f(z)\rangle_x$ is the $Z$ transform of the wave function
free of measurements initially localized on site $x$,
$|\psi_f(z)\rangle_x=\sum_{n=0} ^\infty z^n |\psi_f(n\tau)\rangle_x$ 
with
$|\psi_f(n \tau) \rangle_x = \exp(- i H n \tau)|x \rangle$. 
Eqs. (\ref{eqRed06},\ref{eqRed06aa}) are the quantum counterparts
of the 
classical renewal equation
Eq. $1.2.3$ (or Eq. I.$18$) in
 \cite{Redner,MontrollW} respectively.  
The latter deals with the correspondence between occupation
and first passage time probabilities, while
 we have found the connection between the amplitudes $\phi_n$
and the wave function $|\psi_f\rangle$. 
In that sense we have reduced the problem of  first detection time
to the computation of the $Z$ transform of the solution of the
Schr\"odinger equation. 

\begin{figure}
\centering
\includegraphics[width=0.42\textwidth]{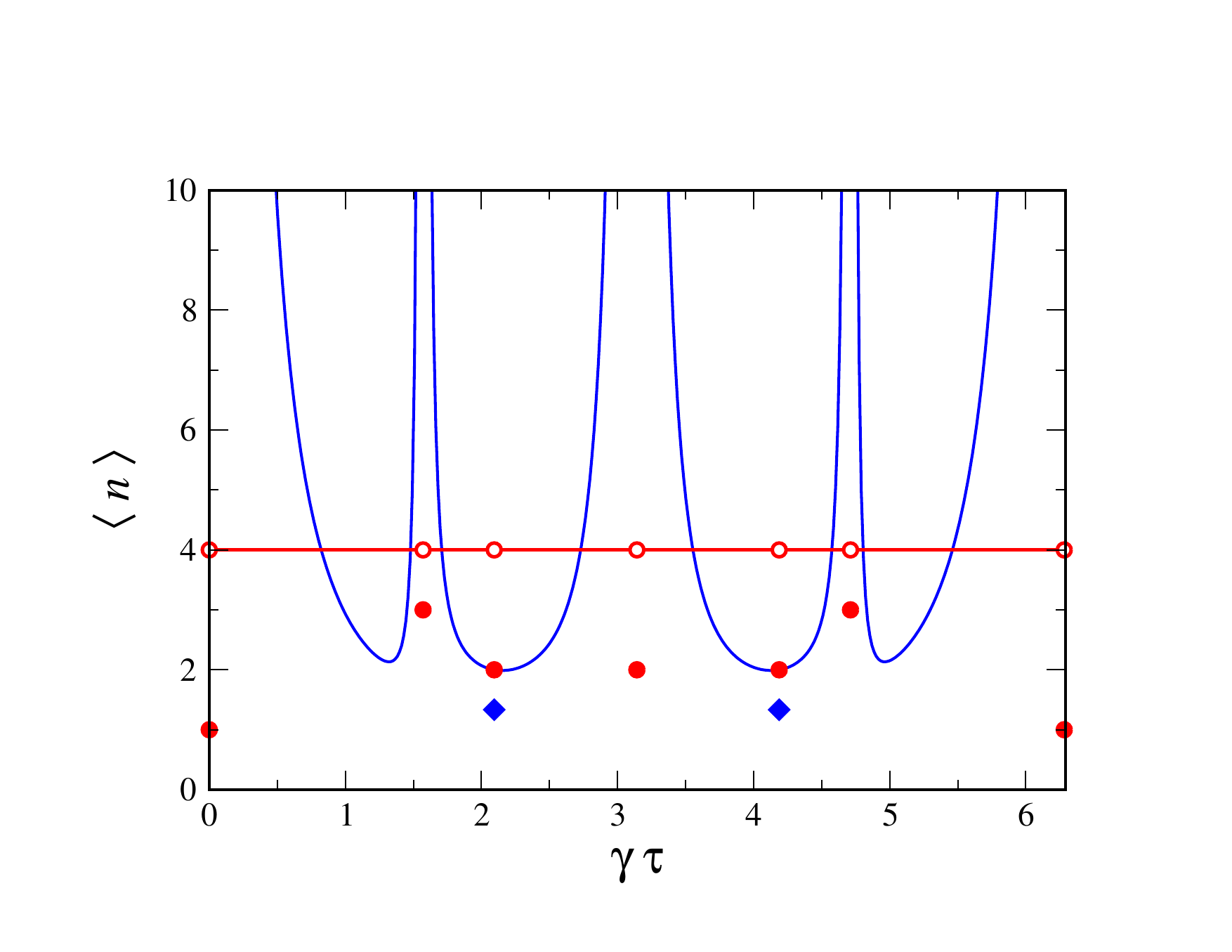}
\caption{
The average first successful detection
 $\langle n \rangle$ versus $\gamma \tau$ for a benzene-like ring
where  the starting point is on $x=3$ (blue) or $x=0$ (red)  and the 
detection is on $x=0$.
}
\label{fig2three}
\end{figure}

{\em Rings.} 
We first consider a tight-binding ring with $L$ sites, namely the
Hamiltonian Eq. (\ref{eq01}) with periodic boundary conditions.
The solution of the  Schr\"odinger equation $|\psi_f\rangle$
is computed with standard methods. 
To find $\phi_n$ we use the inverse
$Z$ transform  (see SM). 
For an $x=0$  initial condition, i.e. $|\psi(0)\rangle=|0\rangle$, where
$x=0$  is also the
location of the detector, we find the following
three results:
(i) The particle is detected with probability $1$ and in this
sense the quantum walk is recurrent.
(ii)  Besides isolated sampling times $\tau$ listed below,
the average number of detection attempts is
\begin{equation}
\langle n \rangle =\left\{
\begin{array}{c c}
{ L + 2 \over 2} & \  \mbox{$L$ is even} \\
\ & \ \\
{ L + 1 \over 2} & \  \mbox{$L$ is odd}.
\end{array}
\right.
\label{eqKRSL}
\end{equation}
This result is remarkable since it is independent of the sampling time $\tau$.
(iii)
Exceptional sampling times $\tau$ are given by the rule
\begin{equation}
\Delta E \tau = 2 \pi n,
\label{eqET}
\end{equation}
where $n$ is a non-negative
integer, and $\Delta E= E_i - E_j >  0$ is the energy
difference between pairs of eigenenergies of the underlying Hamiltonian.
 These exceptional points exhibit non-analytical behaviors,
diverging moments of $n$ and critical slowing down,
as we now discuss.

 For example, for a benzene-like ring of size $L=6$, schematically
shown in the upper part of
 Fig. \ref{figBenzene}, Eq. (\ref{eqKRSL}) gives
$\langle n \rangle=4$. The energy levels
are $\pm  \gamma$ and $\pm 2 \gamma$,
the former are doubly degenerate. 
Using Eq. (\ref{eqRed06}) we find
$\langle n \rangle=1,3,2,2, \cdots$ for the exceptional
points given by Eq. (\ref{eqET})  
$\gamma \tau =0,\pi/2,2 \pi/3, \pi \cdots$ 
which is continued periodically, see Fig.  
\ref{fig2three}.
Physically, the condition
Eq. (\ref{eqET})  implies a partial revival of the wave packet
free of measurement, namely two modes of the system are behaving
identically when strobed at period $\tau$.
When $\tau=0$ we get the expected result, the particle
is detected immediately and then $\langle n \rangle=1$,
which is also found when  $\gamma \tau=2 \pi$, namely
at a full revival period. 

Exceptional sampling times manifest themselves in different ways
depending on the observable and the initial condition.
For example, consider again the average $\langle n \rangle$
for the benzene-like ring but now 
with the initial condition that the particle is
localized at the site  $x=3$ (see Fig. \ref{figBenzene}).
Eq. (\ref{eqRed06aa}) gives, except for the exceptional $\tau$s,
\begin{equation}
\langle n \rangle= 
{27 + 23 \cos(\gamma\tau) + 24\cos(2\gamma\tau) + 9 \cos(3\gamma\tau) - 2 \cos(4\gamma\tau) \over 9 \sin^2(2\gamma\tau)}.
\label{eqDAVID}
\end{equation}
%
%
The result, presented in Fig. 
\ref{fig2three},
 shows that $\langle n \rangle$ diverges 
when $\sin(2 \gamma \tau)\to 0$.
 When $\tau\rightarrow 0$ the measurements
become very frequent and then the 
 probability of measuring the particle approaches
zero and so $\langle n \rangle$ diverges,
 which is the manifestation of the quantum Zeno effect \cite{Misra}.
A similar blowup of $\langle n \rangle$ is observed when
 $\gamma \tau \to 2 \pi$, 
since the wave packet fully revives
on its initial position $x=3$ and so the measurements do not detect the
particle.
Similarly  
$\langle n\rangle$ diverges 
also for $\gamma \tau$ approaching $\pi/2,\pi,3\pi/2$  
due to partial
revivals.  
A far more subtle effect takes place on the special sampling times
$\gamma \tau=2 \pi/3, 4 \pi/3$. There $\langle n \rangle$
 exhibits a discontinuity: on these exceptional points $\langle n \rangle=4/3$
while in their vicinity we find from Eq. (\ref{eqDAVID}) 
$\langle n \rangle \simeq 2$.
Thus the effects of  exceptional points  
on observables are non-trivial.
 We have found several other peculiar
 behaviors for rings \cite{FKB}, 
 but now
we  turn to the case of an unbounded quantum walk, since the
corresponding classical problem is fundamental in stochastic theories,
e.g., it gives the random walk exponents through the long tailed
 first passage PDF 
\cite{Schro}.
Note that Bach, et al. \cite{Bach} treated  the
detection problem  for
 a  discrete time Hadamard  quantum walk,
leading to behaviors different from what we find here.

{\em First detection time for an unbounded quantum walk} described
by the tight binding Hamiltonian 
Eq. (\ref{eq01}) is now investigated.  
For a particle starting at the origin,  
we use 
 $\langle 0 | \psi_f(t) \rangle = J_0 \left( 2 \gamma t \right)$
\cite{Blumen,Krap}
 and  Eq. 
(\ref{eqRed06})
\begin{equation}
\hat{\phi}(z) = 1 - {1 \over \langle 0 |\psi_f(z)\rangle_0} =
1 - {1 \over \sum_{n=0} ^\infty z^n J_0 ( 2 \gamma \tau n)},
\label{eq35Zan}
\end{equation}
where $J_0(x)$ is the Bessel function of the first kind.
Employing 
$J_0 \left( 2 \gamma \tau n \right) \sim  \cos\left( 2 \gamma \tau n - \pi/4\right) / \sqrt{ \pi \gamma \tau n }$  we obtain
the large $n$ limit of $\phi_n$. 
From this asymptotic property of the Bessel function it becomes
clear that the generating function $\langle 0|\psi_f(z)\rangle_0$ does 
not converge
when $z=r\exp(i\theta)$ with $\theta=\pm2 \gamma \tau$ and $r\ge 1$.
Thus, as shown in the SM, when we invert
$\hat{\phi}(z)$ to find $\phi_n$,  we find
two branch   cuts in the  complex $(r,\theta)$ plane. These branch cuts
merge when $2 \gamma \tau$ is an integer multiple
of $\pi$, a mathematical observation which is 
behind the critical behavior we find below.  

\begin{figure}
\centering
\includegraphics[width=0.39\textwidth]{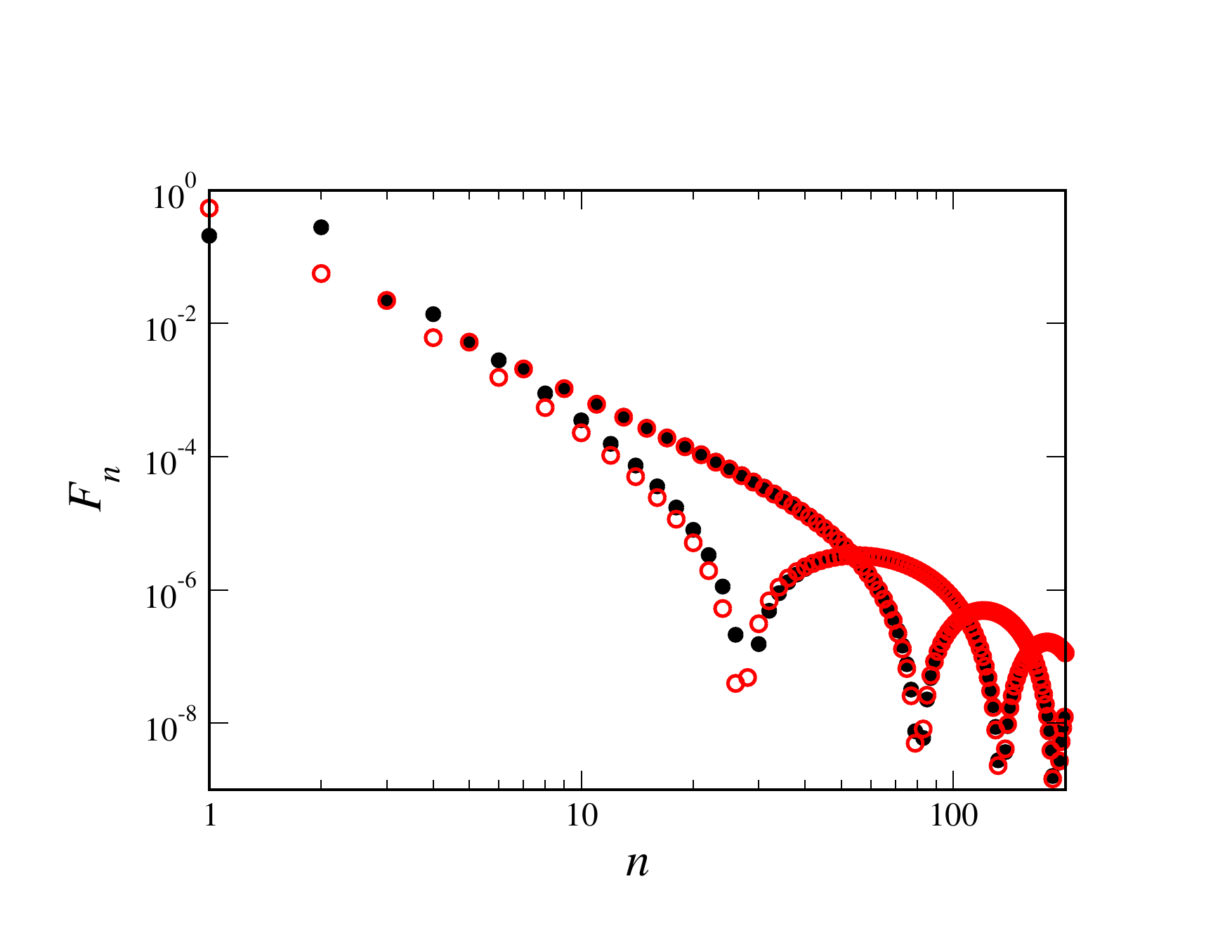}
\caption{
$F_n$  versus $n$ for $\gamma \tau/\pi =0.8/\pi$, for 
a quantum walk on a one dimensional lattice. The asymptotic result
Eq. 
(\ref{eqGen11}) (red open circles)
matches the exact solution already for moderate value of $n$.
The exact solution (full dots)  is obtained using a Taylor series of $\hat{\phi}(z)$
Eq. 
(\ref{eq35Zan}) the $n$-th term yields $\phi_n$ (see details in SM).
}
\label{figNum}
\end{figure}

In the SM we derive another one of our main results: 
 the probability of measuring the quantum walker returning to its
origin for the first time after $n$ attempts  
\begin{equation}
F_n \sim { 4 \gamma \tau \over \pi n^3} \cos^2 \left( 2 \gamma \tau n + {\pi \over 4} \right).
\label{eqGen11}
\end{equation} 
This formula, which is valid for large $n$,
 is the quantum version of the first passage time
problem of a classical one dimensional non-biased
 walker which exhibits the well known  power
law and monotonic
 tail $F_{{\rm classical}} \propto n^{-3/2}$ \cite{Redner,Schro}. 
 The role of sampling time in
the quantum problem is crucial. When $\tau\rightarrow 0$ 
the prefactor of the $n^{-3}$ power law
in Eq. (\ref{eqGen11})
vanishes,
a  manifestation of the quantum  Zeno effect 
(a similar effect is found
for all initial conditions).
Furthermore, the formula predicts that when $\gamma \tau /\pi$ is rational the probability $F_n$ multiplied by $n^3$ is periodic. In contrast
if $\gamma \tau/\pi$ is not rational 
the asymptotic behavior appears irregular (see Fig. 
\ref{figNum}).
In the limit $2 \gamma \tau \rightarrow \pi$
Eq. (\ref{eqGen11})  gives $F_n \sim n^{-3}$
which is a pure power law. However the sampling time
 $2 \gamma \tau=\pi$ is 
exceptional 
 and for this case a detailed calculation reveals
$F_n \sim n^{-3}  /4$ so a factor of $4$ mismatch is found (see SM).
In this sense exceptional points are found also for an infinite system.
 This in turn implies a critical slowing down when $2 \gamma \tau \simeq \pi$
(see SM), a behavior that cannot be anticipated without a detailed calculation.  Physically, the energy band width of a ring of size $L\to \infty$
is $\Delta E= 4 \gamma$ and inserting that in Eq. 
(\ref{eqET}) with $n=1$
we get the exceptional sampling time of the infinite system. 

{\bf Summary.} We have derived 
the long sought after quantum renewal equation,
obtaining the first detection probability of an unbounded
quantum random walk
in one dimension, and finding  unusual non-analytical behavior
even for a small benzene-like ring. Our results are thus the quantum 
version of Schr\"odinger's pioneering  work 
on the classical first passage time
problem from a century ago \cite{Schro}. 
The applications of our main formulas are
vast, since they are not limited to a specific Hamiltonian. 
We note that stroboscopic sampling is very useful in quantum systems,
since this reveals revivals, critical points,
and periodicities, though in principle
the method used in this work could be extended to other measurement protocols. 

{\bf Acknowledgement}  We thank the Israel Science Foundation for funding.


\cleardoublepage
\newpage
\begin{widetext}
\section{Supplementary Material}

\subsection{Generating functions: useful formulas.}

The amplitudes $\phi_n$ are given by the inversion formula
\begin{equation}
\phi_n = { 1 \over  n !} { d^n \over d z^n} \hat{\phi}(z)\Big|_{z=0}
\label{eq22pn}
\end{equation}
or
\begin{equation}
\phi_n = { 1 \over 2 \pi i} \oint_C \hat{\phi}(z) z^{ - n -1} {\rm d} z
\label{eq23}
\end{equation}
where $C$ is a counter-clockwise path that contains the origin and is
entirely within the radius of convergence of $\hat{\phi}(z)$. 
These well-known equations follow from the definition of the generating
function. 
To find $\phi_n$ for a finite $n$, once $\hat{\phi}(z)$ is known,
one may use Mathematica to expand $\hat{\phi}(z)=\sum_{n=1} ^\infty \phi_n z^n$
evaluating the first $n$ terms. 
For the examples we analysed in the text  this method 
is both  practical and fast
for finite though large $n$ (e.g. Fig. 
\ref{figNum} in the text).

The eventual probability of being measured
is related to the generating function $\hat{\phi}(z)$
by
$$ 1-S_\infty=  \sum_{n=1} ^\infty F_n = \sum_{n=1} ^\infty |\phi_n|^2 = $$
\begin{equation}
{1 \over 2 \pi} \int_0 ^{2 \pi} \sum_{k=1} ^\infty \phi_k  e^{ i \theta k}
\sum_{l=1} ^\infty  \phi_l ^{*}  e^{ - i \theta l} {\rm d} \theta=
{1\over 2 \pi} \int_0 ^{2 \pi} |\hat{\phi}(e^{i \theta})|^2 {\rm d} \theta.
\label{eqTot}
\end{equation}
Similarly,
\begin{equation}
\langle n \rangle = \sum_{n=1} ^\infty n F_n =
{1 \over 2 \pi} \int_0 ^{2 \pi} \left[\hat{\phi}\left(e^{i \theta}\right)\right]^{*}
\left(- i {\partial \over \partial \theta} \right)\hat{\phi}(e^{i \theta}) {\rm d} \theta.
\label{eqAve}
\end{equation}
The latter is the average of $n$ only when the particle
is detected with probability one, namely when  $S_\infty=0$.
Thus with the knowledge of the generating function we may compute 
the statistical
properties of  the first detection problem.

\subsection{Benzene-like ring} 

 As we showed in the text, to find $\hat{\phi}(z)$ one must first
find $|\psi_f\rangle$, and as well known, this time dependent solution
of the Schr\"odinger equation is governed by the energy spectrum $\{ E_k \}$ and eigenstates of the Hamiltonian,
together with the initial condition.
 For a benzene-like ring, with initial condition on $x=0$ we
find
\begin{equation}
\hat{\phi}(z) = {{ 1 \over 6} \sum_{k=0} ^5 { 1 \over z^{-1} \exp( i E_k \tau)-1}
\over 
1 + {1 \over 6} \sum_{k=0} ^5 { 1 \over z^{-1} \exp( i E_k \tau) -1}}.
\label{eq31}
\end{equation} 
The nondegenerate energy levels $\{ E_k \}$ $(k=0,...,5)$
 are $-2 \gamma$ and $2 \gamma$ 
while $-\gamma$ and $\gamma$ are doubly degenerate. We  use Eqs.
(\ref{eq23},
\ref{eqTot},
\ref{eqAve}) to derive results presented  in the text, and similarly 
for other initial conditions. The integrals can be evaluated analytically (to be
published) but also numerical integration provides sufficiently accurate
results.  We note that the structure of Eq. (\ref{eq31}) remains for a ring
 of any size $L$,
namely the energy levels of the system determine the generating function completely,
for the case when we start at $x=0$.

Table
\ref{TableExcpetional} summarizes the behavior of $S_\infty$ for the 
benzene-like ring. 
For the initial conditions $x=0$ and $x=3$ we find
the classical behavior $S_\infty =0$ for nearly all sampling rates. However,
on what we call exceptional sampling times $\gamma \tau=0,\pi/2,2\pi/3$ etc
we find non-classical behaviour for the initial condition $x=3$.
Another feature presented  in
the table are  half dark states, namely states where the particle
is found with probability $1/2$ for nearly all sampling points.

\begin{table}[h!]
\begin{tabular}{ c | c | c | c | c | c | c | c | c}
\centering
 $x$ & $0<\gamma \tau < 2 \pi ^{*}$  &$ \gamma\tau=  0$ & $\frac{1}{2}\pi$ & $\frac{2}{3}\pi$ & $\pi$ & $\frac{4}{3}\pi$ & $\frac{3}{2}\pi$ & $2\pi$ \\
 \hline
  0 &  0  & 0 &  0  & 0 &  0  & 0 &  0  & 0\\
  1 & 1/2 & 1 & 5/6 & 1 &  1  & 1 & 5/6 & 1\\
  2 & 1/2 & 1 & 1/2 & 1 & 1/2 & 1 & 1/2 & 1\\
  3 &  0  & 1 & 1/3 & 0 &  1  & 0 & 1/3 & 1\\
\end{tabular}
\caption{
The survival probability $S_\infty$ for a quantum walker
on a benzene-type
ring 
for different localized
 starting points $|\psi(0)\rangle = | x \rangle$.
 Measurements are at $x=0$ and hence
initial conditions on sites $1$ and $2$ are equivalent to initial conditions
on $5$ and $4$ respectively. 
$0<\gamma \tau<2 \pi ^{*}$ implies all values of
$\gamma \tau$ in the interval,
 besides the exceptional points listed on the top row. 
}
\label{TableExcpetional}
\end{table}

\subsection{First detection for unbounded particle.}

 Here we analyse the large $n$ behavior of $\phi_n$ for an unbounded tight binding
quantum walk, where initially the walker is localized at the origin.
The analysis is performed via integration in the complex plane using 
Eqs. 
(\ref{eq35Zan},
\ref{eq23}).
We define
\begin{equation}
I(z) = \sum_{\tilde{n}=1} ^\infty z^{\tilde{n}} J_0 \left( 2 \gamma \tilde{n} \tau \right),
\label{eq39}
\end{equation}
and hence the generating function is 
\begin{equation}
\hat{\phi}(z) = {I(z)\over 1 + I(z)} .
\label{eq40}
\end{equation}
As we shall see, the large $n$ behavior of $\hat{\phi}_n$
 is determined by the singularity structure of $I(z)$, 
which in turn is controlled by the large $\tilde{n}$ terms in the sum Eq.
(\ref{eq39}).
In the large $\tilde{n}$ limit  the asymptotic behavior of the
Bessel function is
\begin{equation}
J_0 \left( 2 \gamma \tau \tilde{n} \right) \sim { \cos\left( 2 \gamma \tau \tilde{n} - \pi/4\right) \over \sqrt{ \pi \gamma \tau \tilde{n} } }.
\label{eq42}
\end{equation} 
Thus, we
define 
\begin{equation}
I_{\gamma \tau} (z) = \sum_{\tilde{n}=1} ^\infty z^{\tilde {n}} 
 { \cos\left( 2 \gamma \tau \tilde{n} - \pi/4\right) \over \sqrt{ \pi \gamma \tau \tilde{n} }}
.
\label{eq43}
\end{equation} 
Then the  large $n$ limit of $\phi_n$ is given by the inversion
of 
\begin{equation}
\hat{\phi}(z) \sim { I_{\gamma \tau} (z) \over 1 + I_{\gamma \tau}(z) }.
\label{eq44}
\end{equation} 

\subsection{Infinite system $\gamma \tau =\pi/2$}

Considering the case $\gamma \tau=\pi/2$, we  find  $I_{\pi/2}(z) = \pi^{-1} \mbox{Li}_{1/2}(-z)$,
where  the polylog $\mbox{Li}_{1/2}(z)$ is defined by
\begin{equation}
\mbox{Li}_{1/2}(z)=\sum_{k=1}^\infty \frac{z^k}{\sqrt{k}}.
\end{equation}
We first investigate the branch cut of this function in the complex plane. 

\subsubsection{On the Polylog}

We want to understand the behavior of the polylog function near the singularity at $z=1$. 
The simplest approach is simply to ask Mathematica. 
\begin{equation}
\mbox{Li}_{1/2}(z)\approx  \sqrt{\frac{\pi}{1-z}} + \zeta(1/2) + \ldots
\end{equation}
where $\zeta$ is the Riemann zeta function. Now, for $z$ complex, $z=1+y\pm i\epsilon$, we get
\begin{equation}
\mbox{Li}_{1/2}(z)\approx  \sqrt{\frac{\pi}{-y\mp i\epsilon}} =  \sqrt{\frac{\pi}{ye^{
\mp i\pi}}}=\pm i\sqrt{\frac{\pi}{y}},
\end{equation}
showing the branch cut discontinuity for $z>1$. A simple way to understand this behavior is to note that the
singularity structure is identical to that of 
\begin{equation}
f(z) \equiv \int_0^\infty \frac{dt}{\sqrt{t}}\,e^{t\ln z} = \Gamma(1/2) (-\ln z)^{-1/2}.
\end{equation}
Thus for $z<1$, this diverges as
\begin{equation}
f(z)\approx \sqrt{\frac{\pi}{1-z}},
\end{equation}
exactly as above. The behavior for $z=1+y\pm i\epsilon$ follows similarly.

Another approach is to calculate the discontinuity across the branch cut.  We start from the integral representation of the polylog:
\begin{align}
\mbox{Li}_{1/2}(z) &= \frac{1}{\Gamma(1/2)}\sum_{k=1}^\infty \int_0^\infty \frac{dt}{\sqrt{t}}\, e^{-kt} z^k \nonumber\\
&=\frac{1}{\sqrt{\pi}} \int_0^\infty \frac{dt}{\sqrt{t}}\, \frac{1}{e^t/z - 1}.
\end{align}
This gives
the discontinuity across the branch cut 
$\textrm{Disc}\left[\mbox{Li}_{1/2}(z)\right] = \mbox{Li}_{1/2}(1 + y + i \epsilon) -\mbox{Li}_{1/2}(1+ y -i \epsilon)$
where $\epsilon \rightarrow 0^{+}$
\begin{align}
\textrm{Disc}[\mbox{Li}_{1/2}(z)] &= \lim_{\epsilon \to 0^+} \frac{1}{\sqrt{\pi}}\int_0^\infty \frac{dt}{\sqrt{t}} \left[\frac{1}{\frac{e^t}{1+y}\left(1-i\frac{\epsilon}{1+y}\right)-1} - \frac{1}{\frac{e^t}{1+y}\left(1+i\frac{\epsilon}{1+y}\right)-1}\right]\nonumber\\
&= \lim_{\tilde{\epsilon} \to 0^+}  \frac{1}{\sqrt{\pi}}\int_0^\infty \frac{dt}{\sqrt{t}} \frac{2i\tilde{\epsilon}}{f^2+\tilde{\epsilon}^2}\nonumber\\
&=2i\sqrt{\pi}\int_0^\infty \frac{dt}{\sqrt{t}}\,\delta(f),
\end{align}
where we have defined $\tilde{\epsilon}\equiv \epsilon e^t/(1+y)^2$ and $f\equiv e^t/(1+y)-1$. For small positive $y$, we can find the zero, $t_*$ of $f$:
\begin{equation}
f(t) \approx \frac{1+t}{1+y}-1 \Rightarrow t_* \approx y.
\end{equation}
Thus, doing the integral over $t$ we get
\begin{equation}
\textrm{Disc}[\mbox{Li}_{1/2}(z)] \approx 2i\sqrt{\frac{\pi}{y}},
\end{equation}
which is consistent with what we have gotten previously.

\subsubsection{Undoing the z Transform for $\gamma \tau =\pi/2$}

We want to calculate $\phi_n$,  using the inverse $z$ transform
of
\begin{equation}
\hat{\phi}(z) = \frac{\mbox{Li}_{1/2}(-z)}{\pi + \mbox{Li}_{1/2}(-z)}.
\end{equation}
One can show that
 there are no poles inside the unit circle, 
hence the dominant contribution arises from the branch cut
of the polylog, namely $\hat{\phi}(z)$
 has a branch cut along the negative $z$-axis.  Now, as mentioned
\begin{equation}
\phi_n = \oint \frac{dz}{2\pi i z^{n+1}} \hat{\phi}(z),
\end{equation}
where the contour encloses the origin and does not cross the branch cut. 
We can deform the contour so that it goes counter-clockwise along the circle at infinity from $\theta=-\pi - i \epsilon$ to $\theta=\pi+i \epsilon$, runs along the branch cut
from $z=-\infty + i\epsilon$, goes around the point at $z=-1$ in an infinitesimal circle and returns along the bottom of the branch cut back out to infinity. The contribution from the circle at infinity vanishes and so we are left with the integral along the branch cut, so that
\begin{equation}
\phi_n = \int_{-\infty}^{-1} \frac{dz}{2\pi i z^{n+1}} \textrm{Disc}[\hat{\phi}(z)],
\end{equation}
where as above the discontinuity is taken from above to below the branch cut. 
 Changing variables to $y=-z-1$, we have
\begin{equation}
\phi_n = \int_0^\infty \frac{dy}{2\pi i}\, (-1)^{n+1} e^{-(n+1)\ln (1+y)} \textrm{Disc}[\hat{\phi}(-1-y)].
\end{equation}
For $n\gg 1$, the exponential factor cuts off the integral for $y\gg 1/n$, and so we can use the small $y$ expansion of the rest of the integrand.  Then 
\begin{align}
\hat{\phi}(-1-y\pm i\epsilon) &= \frac{\mbox{Li}_{1/2}(1+y\mp i\epsilon)}{\pi + \mbox{Li}_{1/2}(1+y\mp i\epsilon)}\nonumber\\
&\approx \frac{\mp i\sqrt{\frac{\pi}{y}}}{\pi  \mp i\sqrt{\frac{\pi}{y}}}\nonumber\\
&\approx \mp i \sqrt{\pi y}.
\end{align}
Putting this all together, we get
\begin{equation}
\phi_n \sim \int_0^{\infty} \frac{dy}{2\pi i} (-1)^{n+1} e^{-ny} \left[\mp 2i \sqrt{\pi y}\right] = (-1)^n \frac{\Gamma(3/2)}{\sqrt{\pi}\,n^{3/2}} =  \frac{(-1)^n}{2n^{3/2}}.
\end{equation}
Hence for the exceptional sampling time 
 $2 \gamma \tau=\pi$ we get $F_n =|\phi_n|^2 \sim n^{-3}/4$ as reported in the text. 
As a test, computing $\phi_{20}$ and $\phi_{40}$ numerically, we have $\phi_{20} (20)^{3/2}=0.470$, $\phi_{40} (40)^{3/2}=0.484$, which is consistent with a $1/n$ convergence to $1/2$.

\subsubsection{The General Calculation}

We now consider the unbounded  quantum walk, still starting on the origin, but now  for
a general sampling rate.  
Define $\alpha = 2 \gamma \tau$ and  the function
\begin{equation}
Q(z,\sigma)=\sqrt{\frac{2}{\pi\alpha}} e^{i\sigma\pi/4} \mbox{Li}_{1/2}(e^{i\sigma\alpha} z)
\end{equation}
in terms of which we define
\begin{equation}
R(z)=\frac{1}{2i}[Q(z,1)-Q(z,-1)].
\end{equation}
We want the coefficient of $z^n$ in the Taylor expansion of 
\begin{equation}
\hat{\phi}(z) = \frac{R(z)}{1+R(z)}.
\end{equation}
Now, $\hat{\phi}(z)$ 
 has two branch cuts extending along the rays $z_\pm=e^{\pm i\alpha}(1+y)$ for $y>1$. Note that these coincide in the case $\alpha=\pi$ considered above. $Q(z,1)$ has a branch cut along $z_-$ but is finite along $z_+$ and vice versa. Using our expansion of the polylog,
we have
\begin{equation}
Q(z_+, -1) \approx \sqrt{\frac{2}{\pi\alpha}} e^{-i\pi/4} \left[\pm i\sqrt{\frac{\pi}{y}}\right]= \pm i\sqrt{\frac{2}{\alpha y}}  e^{-i\pi/4},
\end{equation}
and
\begin{equation}
Q(z_+,1) \approx \sqrt{\frac{2}{\pi\alpha}} e^{i\sigma\pi/4} \mbox{Li}_{1/2}( e^{2i\alpha}) .
\end{equation}
Similarly,
\begin{equation}
Q(z_-, 1) \approx \sqrt{\frac{2}{\pi\alpha}} e^{i\pi/4} \left[\pm i\sqrt{\frac{\pi}{y}}\right]= \pm i\sqrt{\frac{2}{\alpha y}}  e^{i\pi/4},
\end{equation}
and
\begin{equation}
Q(z_-,-1) \approx \sqrt{\frac{2}{\pi\alpha}} e^{-i\sigma\pi/4} \mbox{Li}_{1/2}( e^{-2i\alpha}) .
\end{equation}
The discontinuity along the $z_+$ branch cut is then
\begin{equation}
\textrm{Disc}[\hat{\phi}(z_+)] \approx 4\sqrt{\frac{ \alpha y}{2}} e^{i\pi/4},
\end{equation}
and the discontinuity along the $z_-$ branch is 
\begin{equation}
\textrm{Disc}[\hat{\phi}(z_-)] \approx -4\sqrt{\frac{ \alpha y}{2}} e^{-i\pi/4} .
\end{equation}
We thus get
\begin{align}
\phi_n &\approx \int_0^\infty \frac{dy}{2\pi i}e^{-ny}\left[e^{-ni\alpha}\textrm{Disc}[\hat{\phi}(z_+)] + e^{ni\alpha}\textrm{Disc}[\hat{\phi}(z_-)]\right]\nonumber\\
&= \frac{1}{2\pi i\,n^{3/2}} [\Gamma(3/2)\sqrt{8 \alpha}]\cdot [-2i\sin( n\alpha - \pi/4)] \nonumber\\
&= -\sqrt{\frac{2\alpha}{\pi\, n^3}}\,\sin\!\left(n\alpha-\frac{\pi}{4}\right).
\end{align}
This immediately gives $F_n=|\phi_n|^2$  Eq. 
(\ref{eqGen11})
in the main  text. 
For example, for $\alpha=\pi/5$, the formula predicts $\phi_{40} (40)^{3/2} = \phi_{80} (80)^{3/2}=.447$, whereas in truth
(obtained with a symbolic program by Taylor expansion of exact result
of $\hat{\phi}(z)$ to order $z^n$) it is $0.458$ for $n=40$ and $0.453$ for $n=80$, 
so that the error again falls as $1/n$.  For a non-rational 
value of $\alpha/\pi$, the formula also works. 
 For example, for $\alpha=1/5$, the formula predicts $\phi_{80}=-2.36 \cdot 10^{-4}$,
 whereas the exact answer is $-2.44 \cdot 10^{-4}$. The strange aspect of the formula is 
that the limit of $\phi_n$ for $\alpha\to \pi$ 
does not reproduce the value at $\alpha=\pi$ we found above.
  This is no doubt due to the presence of two branch cuts for $\alpha\neq \pi$, 
whereas there is only one at $\alpha=\pi$.  This implies that the
convergence of the formula for $\alpha$ close to $\pi$ must be a little funny. 
 Thus, for $\alpha=0.95\pi$, we have that at $n=40$ the ratio of prediction to exact is $-2.72$ 
(i.e., even the sign is wrong), whereas for $n=60$ the ratio is $0.705$, for $n=80$, the ratio is 
$1.03$, at $n=100$, the ratio is $0.947$, and at $n=120$, the ratio is $0.968$, so things are converging,
 even though even intermediate $n$'s are way off.  The formula also works for $\alpha>\pi$. 
 Thus, for $\alpha=11\pi/7$, the ratio is $0.987$ at $n=190$.
Similar effect takes place for $\alpha$ which is an integer multiple of 
$\pi$. 
Finally we note that unlike a classical random walk in one dimension, the eventual probability of being detected is not unity,
so even a one-dimensional walk is not recurrent (unless $\tau=0$ which is the trivial case).
We will discuss this effect and many others in a longer publication, but for now present $1-S_\infty$ versus
$\gamma \tau$ in Fig. 
(\ref{figFGInf}).

\end{widetext}
\begin{figure}
\centering
\includegraphics[width=0.39\textwidth]{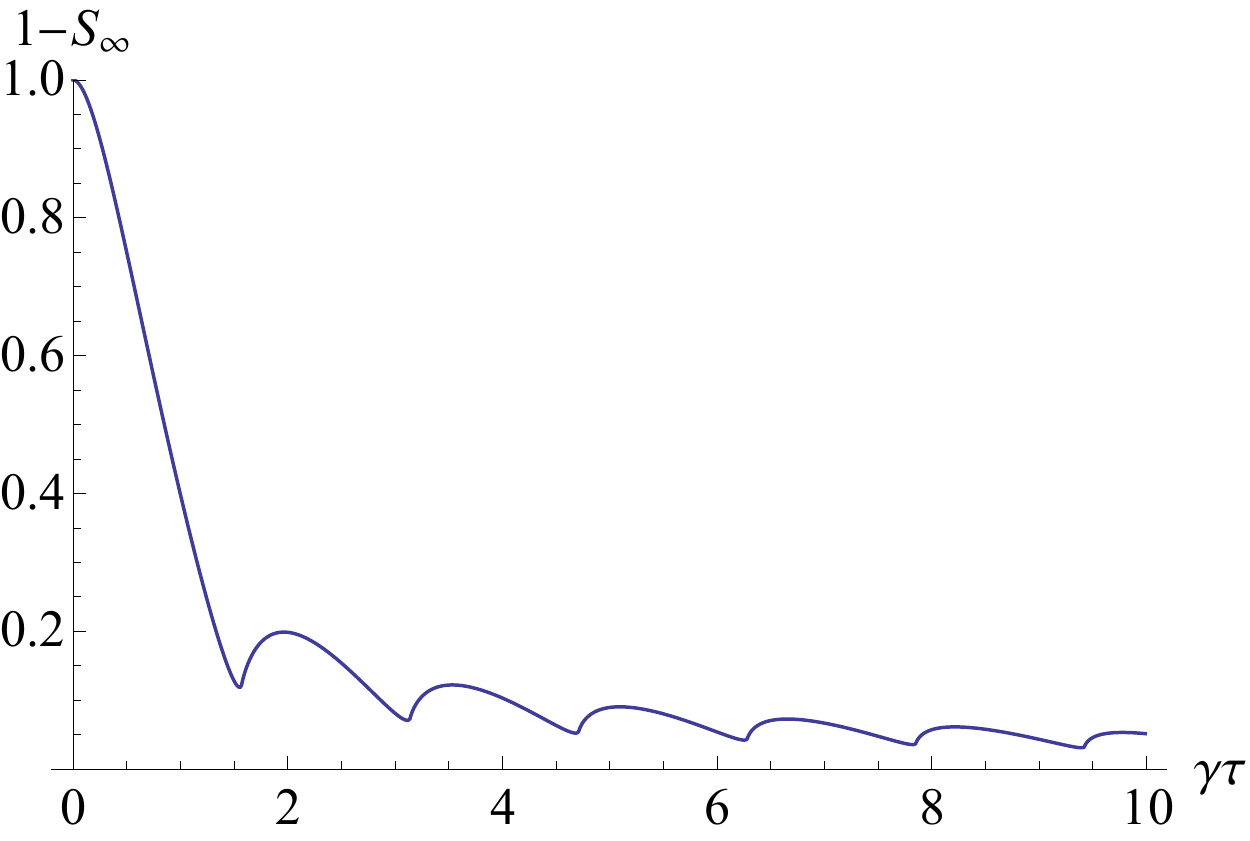}
\caption{
For the one dimensional tight binding quantum walk the
 probability that the particle
is eventually detected $1-S_\infty=\sum_{n=1} ^\infty F_n$ 
versus the sampling rate $\gamma \tau$ exhibits a non-monotonic behavior. 
Unlike the classical random walk
counterpart the quantum  walk is not recurrent,
 unless $\gamma \tau \to 0$ which is the
trivial case.
}
\label{figFGInf}
\end{figure}

\end{document}